\documentclass[superscriptaddress,twocolumn,nofootinbib]{revtex4}
\usepackage{amssymb, bm}
\usepackage{graphicx}

\begin{document}

\title{Determination of Rashba-coupling strength for surface two-dimensional
electron gas in InAs nanowires}

\author{I.A. Kokurin}
\email{kokurinia@math.mrsu.ru} \affiliation{A. F. Ioffe
Physical-Technical Institute, Russian Academy of Sciences, 194021
St. Petersburg, Russia} \affiliation{Institute of Physics and
Chemistry, Mordovia State University, 430005 Saransk, Russia}

\date{\today}

\begin{abstract}
A key concept in the field of semiconductor spintronics is an
electric field control of spins via the spin-orbit coupling (SOC)
and the SOC strength governs efficiency of this control. We propose
a new approach that allows the experimental determination of the
Rashba SOC strength for ballistic InAs nanowires. The energy
spectrum and ballistic transport of carriers through the nanowire
with surface two-dimensional electron gas (2DEG) in a homogeneous
magnetic field are studied. A general formula for the
linear-response one-dimensional ballistic thermopower is derived in
the case of complex subband structure. The ballistic conductance and
the thermopower are shown to reveal specific features due to strong
SOC that allows us to propose a method for the SOC strength
determination.
\end{abstract}


\maketitle

\section{Introduction}
\label{sec:Intro}

Nanowires of narrow gap III-V semiconductors, such as InAs, have
recently attracted significant interest in the field of
nanoelectronics. InAs nanowire is a good candidate for application
in nanodevices such as field effect transistor
(FET)~\cite{Dayeh2007,Chuang2013}. The two-dimensional electron gas
(2DEG) is formed close to the surface of an InAs nanowire due to the
band bending and the Fermi-level pinning~\cite{Hernandez2010} (see
Fig.~\ref{fig1}a). Thus, a one-dimensional (1D) tubular conducting
channel arises near the nanowire surface. Moreover, asymmetric
confinement of the surface 2DEG leads to strong Rashba spin-orbit
coupling (SOC)~\cite{Bychkov1984}. A similar system has been studied
theoretically in more complex model \cite{Jin2010} where the
question of experimental determination of the SOC strength was
discussed. A possibility of SOC strength tuning by back or
surrounding gate~\cite{Liang2012} allows to utilize the nanowires in
spintronics, e.g. as the basic element of spin-FET proposed by Datta
and Das~\cite{Datta1990} or a gate-defined spin-orbit qubit
\cite{Nadj-Perge2010}.

Ballistic transport is preferable for spintronic nanodevices but
implementation of the ballistic transport regime in InAs nanowires
has long hampered due to low carrier mobility that is determined by
the surface roughness scattering. However, it was recently
demonstrated that the ballistic transport can be realized in short
InAs nanowires~\cite{Chuang2013}.

Information about the magnitude of spin-splitting and in turn the
Rashba SOC strength is important for applications. The
antilocalization measurements usually used for experimental
determination of Rashba parameter in
nanowires~\cite{Hernandez2010,Liang2012}. However, such a scheme
does not work in the case of ballistic structures. Therefore there
is a need to develop a new experimental method to determine the
Rashba-coupling strength in ballistic regime.

In present work we study ballistic transport (conductance and
thermopower) through the InAs nanowire and propose a method for
determining the Rashba SOC strength from transport measurements.

\section{Model and spectral problem}
\label{sec:2}

\begin{figure*}
\includegraphics[width=1.3\columnwidth]{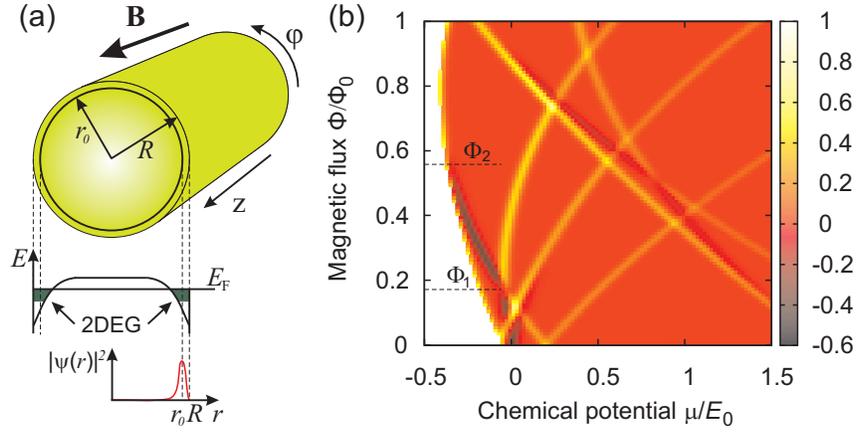}
\caption{\label{fig1} (Color online) (a) Sketch of the InAs nanowire
with radius $R$ placed in homogeneous longitudinal magnetic field.
The band bending which leads to surface 2DEG formation is depicted.
Carriers are concentrated in thin cylindrical layer of radius $r_0$.
The radial distribution (squared amplitude of the wave function
$|\psi|^2$) is schematically shown. (b) The contour plot of the
thermopower $S$ (in units of $k_\mathrm{B}/e$) vs the magnetic flux
and the chemical potential, $\Lambda=0.8$, $T=0.01E_0$. The specific
$\Phi$-values are depicted.}
\end{figure*}

In order to describe spectral properties of the surface 2DEG in InAs
nanowire, we use a simple model of the electron moving on a
cylindrical surface of radius $r_0$ \footnote{The first radial
subband can be taken into account only and the problem could be
considered as the effectively quasi-two-dimensional one. It follows
from simple estimates that gives about 100 meV for the
size-quantization energy (at the well width about 10 nm and the
effective mass $0.026m_0$) whereas the Fermi energy is about 150 meV
as well as from the strict consistent solution of Poisson and
Schr\"{o}dinger equations~\cite{Hernandez2010,Bringer2011}.}. Note
that the surface 2DEG radius $r_0$ does not coincide with nanowire
radius $R$ (see Fig.~\ref{fig1}a). For InAs nanowire with $R=50$ nm
the highest electron density is at $r_0=42$ nm as was shown in
Ref.~\citenum{Bringer2011} by means of consistent solution of
Poisson and Schr\"{o}dinger equations. Moreover, applied magnetic
field will change the surface 2DEG radius. But in our calculations
we suppose the magnetic field to be weak enough and neglect such a
dependence.

The Rashba SOC Hamiltonian on the cylindrical surface that was first
proposed by Magarill et al.~\cite{Magarill1996} is given by
\begin{equation}
\label{Rashba} H_{so}=\frac{\alpha}{\hbar}(\sigma_z
p_\varphi-\sigma_\varphi p_z),
\end{equation}
where $\alpha$ is the Rashba SOC parameter,
$p_z=-i\hbar\partial/\partial z$,
$p_\varphi=-i(\hbar/r_0)\partial/\partial\varphi$, and
$\sigma_\varphi=-\sin\varphi\sigma_x+\cos\varphi\sigma_y$ with
$\sigma_i$ ($i=x,y,z$) being the usual Cartesian Pauli matrices.
Other authors (see for instance Ref.~\citenum{Trushin2007}) used SOC
Hamiltonian that differs from corresponding one of
Ref.~\citenum{Magarill1996} in sign \footnote{In
Refs.~\citenum{Magarill1996,Trushin2007} the model Hamiltonian
(\ref{Rashba}) was considered in other context and the direction of
band banding was not exactly taken into account. However, the
relationship between the direction of the normal to the cylindrical
surface (inside or outside) and the Rashba-coupling sign was
discussed \cite{Magarill1996}. The above inconsistency in sign of
Eq.~(\ref{Rashba}) is apparently due to the domination in the
literature of the following form of Rashba SOC operator
\cite{Bychkov1984},
$H_{so}=(\alpha/\hbar){\bm\sigma}\times\mathbf{p}\cdot\mathbf{n}$,
with $\mathbf{n}$ being the unit vector of the normal to 2DEG (not
the unit vector of built-in or external electric field) that,
however, does not cause misunderstandings in the case of flat
2DEG-structure.}. We choose it in such a form because the direction
of the normal to 2DEG (in our case it is the inside radial
direction) has to coincide with the direction of electric field that
leads to the band bending and Rashba-effect \cite{Winkler2003}.

The one-electron Hamiltonian on the cylindrical surface in the
presence of Rashba SOC and Zeeman splitting in the uniform
longitudinal magnetic field ($\mathbf{B}||z$) is given by
\begin{equation}
\label{Hamiltonian}
H=\frac{\Pi_z^2+\Pi_\varphi^2}{2m^*}+\frac{\alpha}{\hbar}
(\sigma_z\Pi_\varphi-\sigma_\varphi
\Pi_z)+\frac{1}{2}g^*\mu_\mathrm{B}B\sigma_z,
\end{equation}
where $m^*$, $g^*$, $\mu_\mathrm{B} =|e|\hbar/2m_0c$ are effective
mass, g-factor and Bohr magneton, respectively,
$\bm{\Pi}=\mathbf{p}-\frac{e}{c}\mathbf{A}$ is the kinetic momentum
with $\mathbf{A}$ being the vector potential of the magnetic field
($A_\varphi=Br_0/2$, $A_z=0$) \footnote{It should be noted the
vector potential that enters Eq.~(\ref{Hamiltonian}) is no longer
real vector potential and equation
$\mathbf{B}=\nabla\times\mathbf{A}$ fails for it. The real vector
potential that yields the correct $\mathbf{B}$ is $r$-dependent and
it enters the initial three-dimensional Hamiltonian that has to be
averaged over ground state of transverse motion (radial direction)
in order to obtain the Hamiltonian with lower dimension.}.

Since the Hamiltonian (\ref{Hamiltonian}) commutes with $p_z$
(translational invariance) and with the operator of $z$-projection
of total angular momentum
$j_z=-i\hbar\partial/\partial\varphi+(\hbar/2)\sigma_z$ (rotational
invariance) we will look for the eigenstates in the following form
\begin{equation}
\label{eigenstates} \Psi(\varphi, z)=\frac{e^{ikz}}{\sqrt{2\pi L}}
\left(\matrix{e^{i(j-1/2)\varphi}C_{jk}^{(m)}\cr
e^{i(j+1/2)\varphi}D_{jk}^{(m)}\cr}\right),
\end{equation}
where $L$ is the nanowire length, $\hbar k$ is the longitudinal
momentum, and the spinor components $C_{jk}^{(m)}$ and
$D_{jk}^{(m)}$ are in general $k$-dependent due to SOI.

\begin{figure*}
\includegraphics[width=1.3\columnwidth]{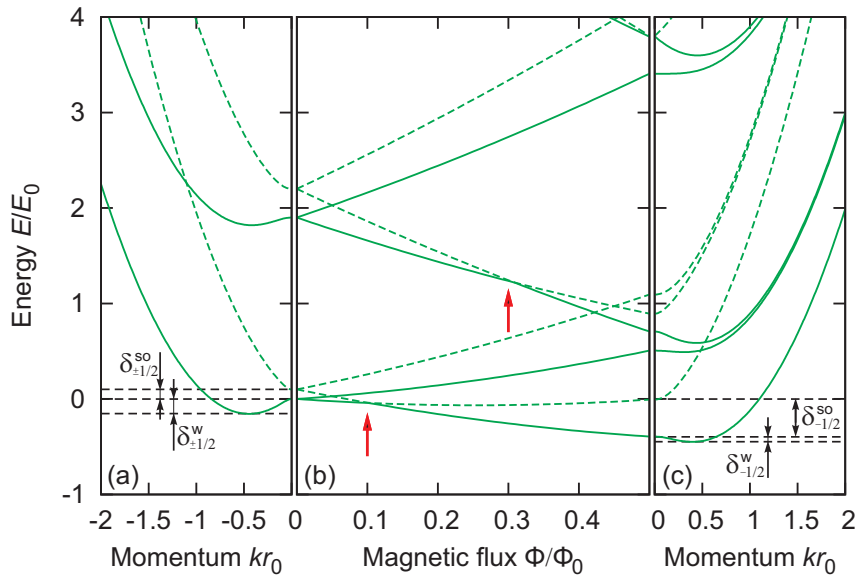}
\caption{\label{fig2} (Color online) The influence of magnetic flux
on energy spectrum of the surface 2DEG, $\Lambda=0.9$. Full (dashed)
lines denote subbands with $m=-1$ ($m=+1$). (a), (c) The subband
spectrum: (a) $\Phi=0$; (c) $\Phi=0.5\Phi_0$. The magnitudes
$\delta_j^{so}$ and $\delta_j^w$ are depicted for the lowest
subbands. (b) The evolution of 1D-subbands with magnetic flux
increasing at $k=0$. Arrows indicate the magnetic flux values at
which the subbands $(j,-1)$ and $(j,+1)$ coincide at $k=0$ (there is
no SO-gap).}
\end{figure*}

The solution of Schr\"{o}dinger equation for the Hamiltonian
(\ref{Hamiltonian}) with the obvious periodicity condition
$\Psi(\varphi+2\pi,z)=\Psi(\varphi,z)$ yields for the energy
spectrum
\begin{eqnarray} \label{spectrum} \nonumber
\frac{E_{jm}(k)}{E_0}&=&(kr_0)^2+\left(j+\Phi/\Phi_0\right)^2+1/4-\Lambda/2\\
&+&m\sqrt{(\Lambda kr_0)^2+[\Delta-(1-\Lambda)(j+\Phi/\Phi_0)]^2},
\end{eqnarray}
where $E_0=\hbar^2/2m^*r^2_0$ is the character energy scale in the
problem, $\Lambda =r_0/l_{so}$ is the dimensionless SOC parameter
with $l_{so}=\hbar^2/2m^*\alpha$ being the character SOC length,
$j=\pm 1/2, \pm 3/2,...$ is the $z$-projection of total angular
momentum, $m=\pm 1$ labels two branches of the spin-split energy
spectrum, $\Phi$ is the magnetic flux $\pi r^2_0B$ through a section
of the surface 2DEG, $\Phi_0=2\pi\hbar c/|e|$ is the flux quantum,
and $2\Delta=g^*\mu_\mathrm{B}B/E_0$ is the dimensionless Zeeman
splitting.

The normalized eigenspinors in (\ref{eigenstates}) are given by
\begin{equation}
C_{jk}^{(+)}=D_{jk}^{(-)}=\cos\left(\frac{\theta_{jk}}{2}\right),\;\;
D_{jk}^{(+)}=C_{jk}^{(-)}=i\sin\left(\frac{\theta_{jk}}{2}\right),
\end{equation}
where $\tan\theta_{jk}=\Lambda
kr_0/[\Delta-(1-\Lambda)(j+\Phi/\Phi_0)]$.

The energy spectrum and its evolution with magnetic field is
depicted in Fig.~\ref{fig2}. At zero magnetic field the spectrum
(\ref{spectrum}) corresponds to the result of
Ref.~\citenum{Magarill1996}. The applied magnetic field lifts the
degeneracy on the sign of $j$. For strong SOC there are W-shape
subbands with $m=-1$ (there are two minima and one maximum), which
appearance depends on relation between $\Lambda$ and $\Phi$. The
W-like shape exists in $(j,-1)$-subband if the following condition
is fulfilled, $\Lambda^2/2-|\Delta-(1-\Lambda)(j+\Phi/\Phi_0)|>0$.

The so-called spin-orbit (SO) gap~\cite{Quay2010} occurs in
above-mentioned subbands. It is the energy distance between
$(j,-1)$-subband maximum and $(j,+1)$-subband minimum. The width of
SO-gap (in $E_0$ units) for $(j,m=\pm 1)$-subbands is
\begin{equation}
\label{delta_so}\delta^{so}_j=2|\Delta-(1-\Lambda)(j+\Phi/\Phi_0)|
\end{equation}
and can vanish at specific $\Phi$-values (Fig.~\ref{fig2}). The
observation of SO-gap was experimentally performed in quantum wires
that defined in two-dimensional hole gas~\cite{Quay2010}.

The energy distance between the two minima and one maximum in
W-shape subband is given by
\begin{equation}
\label{delta_w}
\delta^{w}_j=\Lambda^2/2-|\Delta-(1-\Lambda)(j+\Phi/\Phi_0)|
\end{equation}
and at specific $\Phi$ tends to zero leading to the flat subband
shape at $k=0$. We use these properties below for determination of
SOC strength. The magnitudes $\delta_j^{so}$ and $\delta_j^w$ are
depicted in Fig.~\ref{fig2}.

The SO-gap can occur even in zero magnetic field unlike the case of
planar quantum wires of Ref.~\citenum{Quay2010}. It take place under
the following condition, $\Lambda^2>2|(1-\Lambda)j|$. If there is
SO-gap then its width varies linearly with the SOC strength. Our
estimations show that the W-like subband shape can take place even
at $\Lambda\sim 0.05$, but in this case the $\Phi$-range of the
W-shape structure existence is very narrow and the energy distance
between extrema $\delta^w_j$ is small as well, so that the low
temperature about 10 mK smoothes all relating effects. Thus, SOC can
be assumed to be strong at $\Lambda\gtrsim 0.5$.

\section{Ballistic conductance and thermopower}
\label{sec:3}

Let us consider ballistic transport in the above system. There are
some differences between the well-known ballistic transport through
the 1D system with a parabolic dispersion and the transport in a
system with a complex spectrum. At first, we write the general
formulae for finite-temperature ballistic conductance and
thermopower (Seebeck coefficient) of the quasi-1D system in the case
when subbands have an arbitrary number of extrema.

Now we consider a system that consists of two electron reservoirs
with chemical potentials $\mu_{L(R)}$ and temperatures $T_{L(R)}$
(the temperature is in the energy units) connected by a nanowire. We
assume the transport through the nanowire to be ballistic, i.e.
there is no scattering inside the system and the scattering at the
nanowire-reservoir contact region is negligible. Therefore a
transmission coefficient $T(E)$ within any subband is
energy-independent and equals to unity (i.e. there is the perfect
transmission and no mode mixing,
$T_{j'm',jm}(E)=\delta_{j'j}\delta_{m'm}$). If dc bias $V$ and
temperature difference $\Delta T$ are applied between reservoirs, so
that $\mu_L=\mu_R-eV$ and $T_L= T_R-\Delta T$ ($\Delta T>0$), then
the net current $I$ is given by (see for instance
Ref.~\citenum{Datta1995})
\begin{eqnarray}
\label{current} \nonumber
I=\frac{e}{2\pi}\sum_{j,m}\int_{-\infty}^\infty dk
v_{jmk}[\theta(v_{jmk})f(E,\mu_L,T_L)\\
+\theta(-v_{jmk})f(E,\mu_R,T_R)],
\end{eqnarray}
where $v_{jmk}=\partial E_{jm}(k)/\partial (\hbar k)$ is the
$(j,m)$-subband electron velocity, $\theta(x)$ is the Heaviside unit
step function, $f(\varepsilon,\mu,T)=\{1+\exp
[(\varepsilon-\mu)/T]\}^{-1}$ is the Fermi distribution function.
Here $\theta(\pm v_{jmk})$ ensures that the contribution to the
current is given by electrons moving from the left (right) reservoir
to the right (left) one only. It should be noted, that within
Landauer-B\"{u}ttiker transport framework reservoirs enter the
transport equations through the chemical potential only (see for
instance Refs.~\citenum{Beenakker1991,Imry1999} and references
therein) and there is no influence of energy dispersion in
reservoirs on 1D transport.

After transition to $E$-integration the formula for the net current
can be rewritten in the form
\begin{eqnarray}
\label{net_current} \nonumber I=\frac{e}{2\pi\hbar}\sum_{j,m}\left[
\int_\infty^{E_{jm}^{(1)}}dEf(E,\mu_R,T_R)
\right.\\
\left.
+\int_{E_{jm}^{(1)}}^{E_{jm}^{(2)}}dEf(E,\mu_L,T_L)+...+\int_{E_{jm}^{(N_{jm})}}^\infty
dEf(E,\mu_L,T_L)\right],
\end{eqnarray}
where $E_{jm}^{(n)}$ is the energy value at $n$-th local extremum of
$(j,m)$-subband and total number of extrema in the subband $(j,m)$
is $N_{jm}$. We have one or three extrema as was discussed earlier.

If applied bias voltage $V$ and temperature difference $\Delta T$
are assumed to be low, $\Delta\mu=-eV\ll \mu_{L(R)}$, $\Delta T\ll
T_{L(R)}$ (linear-response regime), then the conductance and
thermopower

\begin{equation}
\label{G_and_S} G=\left(\frac{eI}{\Delta\mu}\right)_{\Delta
T=0},\qquad S=\frac{k_\mathrm{B}}{e}\left(\frac{\Delta\mu}{\Delta
T}\right)_{I=0}
\end{equation}
can be found from (\ref{net_current}). Here $k_\mathrm{B}$ is the
Boltzmann constant.

One can obtain from (\ref{net_current}), (\ref{G_and_S}) Pershin et
al. result for conductance \cite{Pershin2004}
\begin{equation}
\label{conductance} G=\frac{G_0}{2}\sum_{jm}\sum_n\beta_{jm}^{(n)}
f(E_{jm}^{(n)},\mu,T).
\end{equation}
Here $G_0=e^2/\pi\hbar$ is the conductance quantum (for
spin-degenerate case), and $\beta_{jm}^{(n)}=+1$ if $n$-th extremum
of $(j,m)$-subband is the minimum point but $\beta_{jm}^{(n)}=-1$ if
$n$-th extremum of $(j,m)$-subband is the maximum one. The sum in
(\ref{conductance}) is over all extremal points of all subbands.

For the thermopower $S$ we find from (\ref{net_current}) and
(\ref{G_and_S}) the following equation
\begin{equation}
\label{thermopower} S=\frac{k_\mathrm{B}}{e}
\frac{\sum_{jm}\left[\ln 2+\sum_n\beta_{jm}^{(n)}
F\left(\frac{E_{jm}^{(n)}-\mu}{2T}\right)\right]}{\sum_{jm}\sum_n\beta_{jm}^{(n)}
f(E_{jm}^{(n)},\mu,T)},
\end{equation}
where the function $F(x)=\ln(\cosh x)-x\tanh x$ has the following
properties: $F(-x)=F(x)$, $F(\pm\infty)=-\ln 2$, and $F(0)=0$. The
function $F[(E-\mu)/2T]$ as a function of the chemical potential
$\mu$ represents a narrow symmetric peak with a width about several
$T$.

\begin{figure*}
\includegraphics[width=1.3\columnwidth]{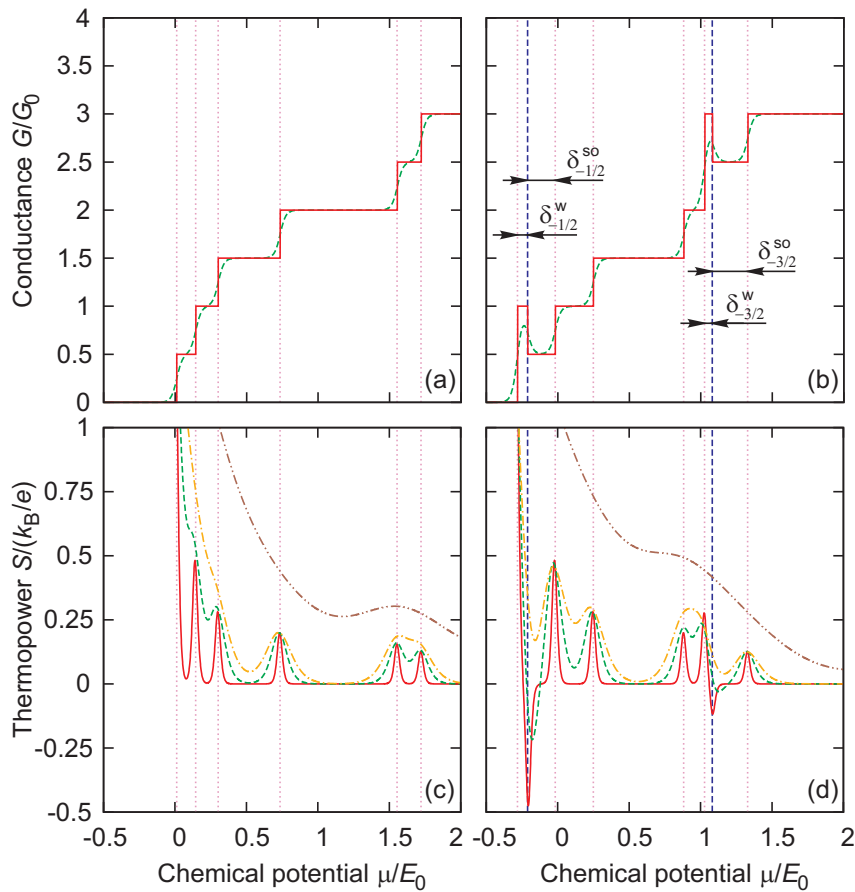}
\caption{\label{fig3} (Color online) Conductance (a), (b) and
thermopower (c), (d) as a function of the chemical potential. (a),
(c) Weak SOC, $\Lambda=0.15$, $\Phi=0.28\Phi_0$; (b), (d) Strong
SOC, $\Lambda=0.78$, $\Phi=0.34\Phi_0$; (a), (b) Full line, $T = 0$;
dashed line, $T=0.02E_0$; (c), (d) Full line, $T=0.01E_0$; dashed
line, $T=0.03E_0$; dot-dashed line, $T=0.05E_0$; double-dot-dashed
line, $T=0.2E_0$. The vertical dashed (dotted) lines denote the
position of the energy maxima (minima). The magnitudes
$\delta_j^{so}$ and $\delta_j^w$ are depicted for the case of strong
SOC.}
\end{figure*}

In accordance with Eqs.~(\ref{conductance}) and (\ref{thermopower}),
positions of conductance up-(down-)steps and thermopower peaks
(dips) correspond to coincidence of the chemical potential with the
minimum (maximum) in the energy spectrum. The dependencies of $G$
and $S$ on the chemical potential $\mu$, calculated by means of
Eqs.~(\ref{conductance}) and (\ref{thermopower}), are depicted in
Fig.~\ref{fig3}. One can see the usual step-like conductance
increasing with $\mu$-increasing and regular thermopower
oscillations \cite{Streda1989,Proetto1991} in the case of weak SOC
(in this case the 1D-subbands have near-parabolic shape and single
minimum is in each subband). Each spin-split subband contributes
$+G_0/2$ to the total conductance. In the case of simple subband
dispersion one can immediately obtain from (\ref{thermopower}) the
Streda result \cite{Streda1989} that determines the thermopower
magnitude at the maxima
\begin{equation}
S_i^\mathrm{max}=\frac{k_\mathrm{B}}{e}\frac{\ln 2}{i+1/2},
\end{equation}
where $i$ is a peak number or a number of occupied subbands. This
equation fulfills at low temperatures when thermopower peaks are
sufficiently narrow and they do not overlap each other.

In the opposite case of strong SOC specific features appear:
conductance down-steps ($-G_0/2$-steps) and negative thermopower
dips. These features correspond to coincidence of the chemical
potential with the maximum in 1D-subband. The presence of successive
$+G_0$-step and $-G_0/2$-step in $G(\mu)$-dependence is the
contribution of W-shape subband. The width of corresponding
$\sqcap$-like plateau is obviously equal to $\delta_j^w$. The SO-gap
$\delta^{so}_j$ is equal to the corresponding $\sqcup$-like plateau
width in conductance and the distance between sequential dip and
peak in thermopower. At strong SOC and $B=0$ (when SO-gap exists)
the dependence $G(\mu)$ will have $+2G_0$, $+G_0$ and $-G_0$ steps
due to the two-fold $j$-degeneracy.

As the temperature increases one can see the smearing of conductance
steps, since electrons coming in from reservoirs no longer have a
sharp steplike energy distribution. The down-step disappears at the
temperature about $\mathrm{min}(\delta_j^{so},\delta_j^w)$. At the
same temperature the corresponding negative thermopower dip vanishes
that is due to the overlapping with a neighboring peak. Thus, low
temperatures are necessary for the negative thermopower observation.

The applied magnetic field changes a position of subband extrema
(and even the number of extrema) and these changes are seen in
transport characteristics. A contour plot of the thermopower vs the
magnetic flux and the chemical potential is shown in
Fig.~\ref{fig1}b. This figure will be discussed below in application
to the determination of the SOC strength parameter.

\section{Determination of Rashba-coupling strength parameter}
\label{sec:4}

Now we discuss the approach for determination of the SOC strength
parameter $\alpha$ from transport measurements that based on
detection of mentioned spectral features evanescence with changing
magnetic field. We do not describe any experimental setups here. The
principal requirement to experimental setup is the possibility of
independent control of Rashba-coupling parameter (that has to be
extracted from meausrements) and Fermi-level position. Metallic
gates or optical excitation can be utilized for the chemical
potential tuning.

In Fig.~\ref{fig1}b one can see two features: (i) the thermopower
peak-dip annihilation ($S=0$) and (ii) dip disappearance close to
the peak of higher amplitude ($S>0$) that occur at $\Phi_1$- and
$\Phi_2$-flux, respectively. It corresponds to above-mentioned
spectral features at which Eqs.~(\ref{delta_so}), (\ref{delta_w})
tend to zero. We propose to utilize these features for experimental
determination of the SOC strength parameter. As was discussed
earlier, up-(down-) steps in conductance and peak (dip) in
thermopower take place at the same chemical potentials. In this
sense conductance measurement is equivalent to thermopower
measurement for our purposes, and mentioned features can be detected
by two methods. Moreover conductance measurements can be simpler and
more straightforward from an experimental point of view. The
magnitude $\partial G/\partial\mu$ (or $\partial G/\partial V_g$,
with $V_g$ being the gate voltage), that is proportional to
thermopower in the low-temperature regime in accordance with the
so-called Mott formula \cite{Cutler1969}, can be measured directly.

Thus, it is necessary for our purpose to perform the thermopower or
conductance measurements in external magnetic field. Specific values
of magnetic field, at which the mentioned features occur, can be
extracted from a dependence of $S$ or $\partial G/\partial\mu$ on
the magnetic field and gate voltage. Inserting these values of
magnetic field into Eqs.~(\ref{delta_so}), (\ref{delta_w}) at
$\delta_j^{so}=0$ and $\delta_j^w=0$, one will obtain the system of
two equations relative to $r_0$ and $\alpha$. It is convenient to
solve these equations for ground subband $j=-1/2$ (as in
Fig.~\ref{fig1}b) and using $m^*=0.026m_0$, $g^*=-14.9$ for InAs
(see for instance Ref.~\citenum{Winkler2003}). Thus, the numerical
solution of equations $\delta^{so}_{-1/2}(\Phi_1)=0$ and
$\delta^w_{-1/2}(\Phi_2)=0$ gives the values $\alpha$ and $r_0$.
Although the above system of equations has the high order, but
numerical analysis shows the presence of the single nontrivial
solution.

\section{Discussion and conclusion}
\label{sec:5}

Let us discuss some weaknesses of proposed approach that deal with
both experimental problems and flaws of the used simple model.

(i) The proposed approach requires the low temperature thermopower
or conductance measurements. We have to compare the temperature not
only with character energy scale $E_0$ but the SO-gap as well in
order to one can experimentally discern single peak (dip) in $S$ or
$\partial G/\partial\mu$. These measurements have to be performed at
temperatures about 1K for the nanowire with $r_0=42$ nm ($R=50$ nm).

(ii) It should be noted that we neglect in our calculation a
dependence of the surface 2DEG radius $r_0$ on the applied magnetic
field. At high magnetic field the character length of the radial
wave-function localization will be determined by the cyclotron
motion, but it does not tend to zero and will be limited by the
effective potential of the band bending that prevents the
penetration of the wave-function in central area of the nanowire.
For the nanowire with $R=50$ nm ($r_0=42$ nm at $B=0$)
\cite{Hernandez2010,Bringer2011} the minimal $r_0$-value (in the
limit $B\rightarrow\infty$) is estimated to be about 30 nm. Our
estimations show that at intermediate magnetic field, corresponding
to the flux $\Phi\sim\Phi_0/2$ (at such a field $r_0\sim l_B$ with
$l_B=\sqrt{\hbar c/|e|B}$ being the magnetic length; it corresponds
to $B\sim 0.35$ T for the nanowire with $R=50$ nm), $r_0$ is about
37 nm that differs only by 12 percent from the magnitude of $r_0$ at
$B=0$. Thus, we hope that our assumption ($r_0$ does not depend on
$\mathbf{B}$) holds for the proposed $\alpha$-determination approach
in not too strong magnetic field. Nevertheless, in order to obtain
more precise estimations it is necessary to solve the consistent
problem \cite{Bringer2011} in magnetic field. However, one should
expect, at least, the determined Rashba-parameter to be of the
correct order of magnitude in our approach.

One can use a simpler approach: if the surface 2DEG radius is known,
e.g. from the consistent scheme \cite{Bringer2011}, then only one
equation is necessary. It is convenient to use
$\delta^{so}_{-1/2}(\Phi_1)=0$, because the weaker magnetic field is
required for the SO-gap disappearance than for the flat subband
appearance. Moreover, there is no problem of $r_0(B)$-dependence at
weak magnetic fields in this case.

Our approach for SOC strength determination is based on transport
detection of SO-gap evanescence in magnetic field. However, for the
InAs nanowires the conductance down-steps (SO-gap) were not yet
observed comparative to quantum wire defined in 2D structure with
high mobility~\cite{Quay2010}. Nevertheless, we guess that the
absence of negative steps in recent conductance quantization
experiments \cite{Chuang2013} deals with no any residual scattering,
but it is due to thermal smearing of steps or the absence of W-shape
subbands in the energy spectrum due to small effective SOC strength
that is proportional to $\alpha r_0$. The latter can be due to the
small effective radius of surface 2DEG. In this sense, our approach
is not applicable for nanowires of small radius.

Recently it was shown \cite{Liang2012} that the Rashba-coupling
parameter can be tuned in the range $\alpha=0.5-3\times 10^{-9}$
eV$\cdot$cm for InAs nanowire that corresponds to dimensionless SOC
parameter $\Lambda=0.15-1$ for the nanowire with $r_0=42$ nm. Thus,
the proposed approach of $\alpha$-determination is valid in the half
of $\alpha$-tuning interval (we suppose that our approach works
adequately at $\Lambda\gtrsim 0.5$).

It would be extremely interesting to compare the results obtained by
proposed method and by the standard antilocalization
technique~\cite{Hernandez2010,Liang2012}. For example, at first one
measures Rashba parameter in long wire by the
localization-antilocalization, and after that the measurements of
SOC parameter are performed by the proposed method in shortened
nanowire where ballistic transport takes place.

In conclusion, we have theoretically studied the spectral and
transport properties of carriers in InAs nanowire with surface 2DEG
in magnetic field. Ballistic conductance and thermopower are shown
to have some specific features in dependence on the chemical
potential that are due to strong Rashba SOC. These features
disappear at certain magnetic fields. The latter allows us to
propose the approach for the experimental determination of
Rashba-coupling parameter $\alpha$ from thermopower or conductance
measurements. Limitations of the proposed approach applicability are
discussed.

\begin{acknowledgments}
The author is grateful to N.S.~Averkiev, P.A.~Alekseev, P.V.~Petrov
and A.Yu.~Silov for useful discussions. This work was supported by
Russian Ministry of Education and Science (project No.~2665).
\end{acknowledgments}


\begin{thebibliography}{99}
\bibitem{Dayeh2007}
S.~A. Dayeh, D.~P.~R. Aplin, X.~Zhou, P.~K.~L. Yu, E.~T. Yu,
D.~Wang, Small \textbf{3}, 326 (2007).

\bibitem{Chuang2013}
S.~Chuang, Q.~Gao, R.~Kapadia, A.~C. Ford, J.~Guo, A.~Javey, Nano
Lett. \textbf{13}, 555 (2013).

\bibitem{Hernandez2010}
S.~Est\'evez~Hern\'andez, M.~Akabori, K.~Sladek, C.~Volk, S.~Alagha,
H.~Hardtdegen, M.~G. Pala, N.~Demarina, D.~Gr\"utzmacher,
T.~Sch\"apers, Phys. Rev. B \textbf{82}, 235303 (2010).

\bibitem{Bychkov1984}
Y.~A. Bychkov, E.~A. Rashba, JETP Lett. \textbf{39} 78 (1984).

\bibitem{Jin2010}
S.~Jin, J.~Waugh, T.~Matsuura, S.~Faniel, H.~Wu, T.~Koga, Physics
Procedia  \textbf{3}, 1321 (2010).

\bibitem{Liang2012}
D.~Liang, X.~P. Gao, Nano Lett. \textbf{12}, 3263 (2012).

\bibitem{Datta1990}
S.~Datta, B.~Das, Appl. Phys. Lett. \textbf{56}, 665 (1990).

\bibitem{Nadj-Perge2010}
S.~Nadj-Perge, S.~M. Frolov, E.~P. A.~M. Bakkers, L.~P. Kouwenhoven,
Nature \textbf{468}, 1084 (2010).

\bibitem{Bringer2011}
A.~Bringer, T.~Sch\"apers, Phys. Rev. B \textbf{83}, 115305 (2011).

\bibitem{Magarill1996}
L.~I. Magarill, D.~A. Romanov, A.~V. Chaplik, JETP Lett.
\textbf{64}, 460 (1996).

\bibitem{Trushin2007}
M.~Trushin, J.~Schliemann, New J. Phys. \textbf{9}, 346 (2007).

\bibitem{Winkler2003}
R.~Winkler, \textit{Spin-Orbit Coupling Effects in Two-Dimensional
Electron and Hole
  Systems.} (Springer-Verlag, Berlin, 2003).

\bibitem{Quay2010}
C.~H.~L. Quay, T.~L. Hughes, J.~A. Sulpizio, L.~N. Pfeiffer, K.~W.
Baldwin,
  K.~W. West, D.~Goldhaber-Gordon, R.~de~Picciotto, Nature Phys.
  \textbf{6}, 336 (2010).

\bibitem{Datta1995}
S.~Datta, \textit{Electronic Transport in Mesoscopic Systems.}
(Cambridge University Press, Cambridge, 1995).

\bibitem{Beenakker1991}
C.~W.~J. Beenakker, H.~van Houten, Solid State Phys. \textbf{44}, 1
(1991).

\bibitem{Imry1999}
Y.~Imry, R.~Landauer, Rev. Mod. Phys. \textbf{71}, S306 (1999).

\bibitem{Pershin2004}
Y.~V. Pershin, J.~A. Nesteroff, V.~Privman, Phys. Rev. B
\textbf{69}, 121306 (2004).

\bibitem{Streda1989}
P.~Streda, J. Phys.: Condens. Matter \textbf{1}, 1025 (1989).

\bibitem{Proetto1991}
C.~R. Proetto, Phys. Rev. B \textbf{44}, 9096 (1991).

\bibitem{Cutler1969}
M.~Cutler, N.~F. Mott, Phys. Rev. \textbf{181}, 1336 (1969).
\end{thebibliography}

\end{document}